\documentclass[structabstract]{aa}  
\usepackage{graphics,latexsym,amssymb,times,psfig}
\usepackage{txfonts}
\newcommand{\swift}{\emph{Swift~}}

\begin{document}
   \title{On the nature of the extremely fast optical rebrightening of the afterglow of GRB 081029}
\authorrunning{Nardini Marco et al.}
   \subtitle{}

   \author{M. Nardini
          \inst{1,2}, J. Greiner\inst{2}, T. Kr\"uhler\inst{2}, R. Filgas\inst{2},  S. Klose\inst{3}, P. Afonso\inst{4},  C. Clemens\inst{2},   A. Nicuesa Guelbenzu\inst{3}, F. Olivares E.\inst{2}, A. Rau\inst{2},  A.  Rossi\inst{3}, A. Updike\inst{5}, A. K{\"u}pc{\"u} Yolda{\c s}\inst{6},  A. Yolda{\c s}\inst{6}, D. Burlon\inst{2}, J. Elliott\inst{2}  \and D. A. Kann\inst{3} }


   \institute{Universit\`a degli studi di Milano-Bicocca, Piazza della Scienza 3, 20126, Milano, Italy\\
              \email{marco.nardini@unimib.it}
              \and
             Max-Planck-Institut f\"ur extraterrestrische Physik, Giessenbachstrasse, 85748 Garching, Germany  
               \and
             Th\"uringer Landessternwarte Tautenburg, Sternwarte 5, 07778 Tautenburg, Germany
               \and
               American River College, Physics and Astronomy Dpt., 4700 College Oak Drive, Sacramento, CA 95841, USA,
               \and
              Department of Physics and Astronomy, Clemson University, Clemson, SC 29634-0978, USA
              \and
              Institute of Astronomy, University of Cambridge, Madingley Road, CB3 0HA Cambridge, UK               
                \\
  }             

   \date{}

 
  \abstract
   {After the launch of the \swift satellite, the Gamma-Ray Burst (GRB) optical light-curve smoothness paradigm has been questioned thanks to the faster and better sampled optical follow-up, which has unveiled a very complex behaviour. This complexity is triggering the interest of the whole GRB community. The GROND multi-channel imager is used to study optical and near-infrared (NIR) afterglows of  GRBs with unprecedented optical and near-infrared temporal and spectral resolution. The GRB 081029 has a very prominent optical rebrightening event and is an outstanding example of the application of the multi-channel imager to GRB afterglows.
}
   {Here we exploit  the rich GROND  multi-colour follow-up of GRB 081029 combined with  XRT observations to study the nature of late-time  rebrightenings that appear in the optical-NIR light-curves of some GRB afterglows. }
   {We analyse the optical and NIR observations obtained with the seven-channel Gamma-Ray burst Optical and Near-infrared Detector (GROND) at the
2.2 m MPI/ESO telescope and the X-ray data obtained with the XRT telescope on board  the \swift observatory. The multi-wavelength temporal and spectral evolution is discussed in the framework of different physical models.}
   {The extremely steep optical and NIR rebrightening observed in GRB 081029 cannot be explained in the framework of the standard forward shock afterglow model. The absence of a contemporaneous X-ray rebrightening and the evidence of a strong spectral evolution in the optical-NIR bands during the rise suggest  two separate components  that dominate in the early and late-time light-curves, respectively.  The steepness of the optical rise  cannot be explained even in the framework of the alternative scenarios proposed in the literature unless a late-time activity of the central engine is assumed.   }
   {}

   \keywords{gamma-rays burst: individual: GRB 081029 -- techniques: photometric               }

   \maketitle
%

\section{Introduction}
For years after the discovery of the first Gamma-Ray Burst (GRB) afterglow (van Paradijs et al., 1997), the smoothness  of the optical afterglow light-curves has been considered one of the main GRB features (Laursen \& Stanek 2003).
Nowadays, thanks to the rapid follow up with robotic telescopes, it is possible to reconsider this paradigm and several examples of complex optical light-curves are known. The Gamma-Ray burst Optical Near-infrared Detector (GROND) is a seven-band simultaneous  optical-NIR  imager  mounted on the 2.2 m MPI/ESO telescope at La Silla observatory (Greiner et al. 2008). GROND is a unique instrument  to study the optical spectral evolution associated to these complex light-curves  (e.g., GRB 071031 Kr\"uhler et al. 2009 and GRB 080129 Greiner et al. 2009). In this paper, we report on the multi-wavelength observation of GRB 081029.  This GRB is characterised by a very complex light-curve with a strong chromatic temporal evolution. In the optical and near-infrared bands an extremely steep rebrightening, at around 3 ks after the trigger, suddenly interrupts the smooth early-time temporal evolution. Thanks to GROND we were able to observe  this event simultaneously form the optical $g^\prime r^\prime i^\prime z^\prime$ to the near-infrared $JHK_s$ bands. 
 This unprecedented temporal and spectral resolution allows a time-resolved analysis of the colour evolution. The analysis of the XRT light-curve excludes the presence of a similar rebrightening in the X-ray bands, which casts doubts on the common nature of the optical and X-ray afterglow emission. 
 
 The existence of late-time rebrightenings in some  GRB optical afterglows has been known since the dawn of afterglow observations  (e.g., the optical bump of  GRB 970508 (Vietri 1998, Sokolov et al. 1998, Nardini et al. 2006)) and several models have been proposed to account for deviations from a smooth power-law evolution in the optical light-curves (see \S \ref{model}).  Some of them, in the framework of standard external-shock afterglow model,  invoke a discontinuity in the external medium density profile (e.g., Dai \& Wu 2003; Lazzati et al. 2002; Nakar \& Piran 2003) some of them considering possible variations of the micro-physical parameters into the fireball (Kong et al. 2010). In other cases a possible energy injection into the fireball (e.g.,  J\'ohannesson et al. 2006;  Fan \& Piran 2006) or complex jet geometry  is considered (e.g.,  Racusin et al. 2009). In the late prompt model (Ghisellini et al. 2007, 2009; Nardini et a. 2010), a late-time activity of the central engine produces optical and X-ray radiation that is superposed on the standard external-shock  afterglow emission.
 
In \S \ref{obs} we present the available broad-band data set. In \S \ref{lc} and in \S\ref{sed} we describe the complex optical and X-ray light-curves and we analyse the broad-band spectal evolution.  In \S \ref{model} we study the possible origin of the optical rebrightening, discussing the observed temporal and spectral properties of the afterglow of GRB 081029  in the framework of different physical models.

\section{Observations and data reduction}
\label{obs}

\subsection{\swift observations}
On 2008 October 29 at 01:43:56 UT, the  \swift Burst Alert Telescope (BAT) triggered on a long  burst (trigger=332931) (Sakamoto et al. 2008) located at coordinates  ${\rm RA}({\rm J}2000)= 23^{\rm h} 07^{\rm m} 06^{\rm s} , {\rm Dec(J}2000) = -68\degr 10' 43\farcs4$ (Cummings et al. 2008). The BAT mask-weighted light-curve is characterised by a single smooth peak starting around 50\,s before trigger, peaking around 60\,s after the trigger, and ending around 300\,s after trigger. The 15-350 keV duration is $T_{90}=270 \pm 45~$\,s. A simple power-law model provides a good fit of the time-integrated  15--350 keV  spectrum with an index $\alpha=1.43\pm 0.18$ ($\chi^2=50.7/57 dof$). In the same BAT energy range GRB 081029 has  a fluence of $2.1\pm 0.2\times 10^{-6}~ {\rm erg~cm}^{-2}$, and a peak flux of $0.5\pm 0.2\times 10^{-6}~{\rm erg~cm^{-2}s^{-1}}$ (Cummings et al. 2008). 
  \\
Owing to observing constraints, the spacecraft could not  immediately slew to the position of the burst.  The X-ray (XRT; Burrows et al. 2005) and UV/Optical Telescope (UVOT; Roming et al. 2005) did not start to observe
the field of GRB 081029 until 2.7 ks after the trigger. XRT started observing in photon counting (PC) mode and found an uncatalogued  source inside the BAT error box located at ${\rm RA}({\rm J}2000)= 23^{\rm h} 07^{\rm m} 05.51^{\rm s} , {\rm Dec(J}2000) = -68\degr 09' 21\farcs9$ (enhanced position obtained combining 2.6 ks of  XRT data and 3 UVOT images; Goad et al. 2008).

\subsection{Optical and NIR observations}
\label{grondobs}
The detection of the optical afterglow of GRB 081029 was first  reported by Rykoff  (2008), who observed an uncatalogued fading source inside the XRT error box with the ROTSE-IIIc telescope. Further observations were reported by Clemens et al. (2008) (GROND), by Covino et al. (2008a) (REM), West et al. (2008) (PROMPT telescopes), and Cobb (2008) (ANDICAM). A redshift of   $z=3.8479\pm 0.0002$ was first reported by D'Elia et al. (2008) from an observation with VLT/UVES and then confirmed by Cucchiara et al. (2008) with GEMINI/GMOS.

\subsubsection{GROND observations and data analysis}
GROND started observing the field of GRB 081029 at about 1:52 UT on 2008 October 29 ($\sim$ 8 minutes after the gamma-ray trigger) and detected the variable source reported by   Rykoff  (2008) in all seven available optical and NIR bands. GROND kept observing GRB 081029  as long as it was visible from La Silla (until about 7:35 UT). Further multi-band observations were taken one, two, three and five days after the trigger. The GROND  optical and NIR image reduction and photometry were
performed using standard IRAF tasks (Tody 1993) similar to the
procedure described  in Kr\"uhler et al. (2008b). A general
model for the point--spread function (PSF) of each image was
constructed using bright field stars, and it was then fitted to the afterglow. Optical photometric calibration was performed relative to the
magnitudes of five secondary standards in the GRB field. During photometric conditions, a primary SDSS standard
field (Smith et al. 2002) was observed within a few minutes
of observing  the GRB field in the first night. The obtained zeropoints were
corrected for atmospheric extinction and used to calibrate stars
in the GRB field. The apparent magnitudes of the afterglow were
measured with respect to the secondary standards. The absolute calibration of the $J H K_{\rm s}$ bands was obtained
with respect to magnitudes of the Two Micron All Sky Survey
(2MASS) stars within the GRB field  (Skrutskie et al. 2006).

\subsubsection{Host galaxy search}
In order to verify the possible presence of a bright ($R_{AB}\sim25$ mag) host galaxy
associated with the burst, we observed the field of GRB~081029 with the
ESO New Technology Telescope (NTT) equipped with EFOSC. 
Observations started at 02:13:37.74 UTC of 2010 November 3, 736 days ($\sim 6.4\times 10^7$\,s) after the burst. Mid-time of the observations is 2010 November 3, 02:46:23 UTC.
A series of
images in the $R$-band filter with a total exposure time of 0.8\,h was
obtained under photometric conditions with a seeing between 0\farcs6 and
0\farcs8. No source is detected down to a $3\sigma$ limiting magnitude of
$R_{AB} > 25.8$ mag, which has been derived by tying the R band magnitudes
of field stars to their GROND $r^\prime$ and $i^\prime$ photometry.\footnote{http://www.sdss.org/dr7/algorithms/sdssUBVRITransform.html \#Lupton2005}

\subsection{X--ray data reduction and spectral analysis}
\label{xspectrum}
We analysed the XRT data of GRB 081029 with the {\it Swift} 
software package  distributed with 
HEASOFT ({\it v6.8}).  The XRT data were reprocessed  with the  \texttt{XRTPIPELINE} 
tool\footnote{Part of the XRT software,
  distributed with the HEASOFT package:  {\tt http://heasarc.gsfc.nasa.gov/heasoft/}}.  
 The entire XRT follow-up of GRB 081029 was performed in PC mode and, since  the 0.3-10 keV observed count-rate never exceeded 0.5 counts $s^{-1}$, no pile-up correction was required (Moretti et al. 2005; Romano et al. 2006; Vaughan et al. 2006). The extraction was 
in  circular regions with typical widths of 25 and 20 pixels depending on the count rate, as discussed 
in Evans et al. (2009). The spectra were extracted  with the standard grade. Background spectra were extracted in  regions of the same size far from the source. 
For all   spectra we created ancillary response files (ARF) with the \texttt{xrtmkarf} 
tool and used the calibration database updated to January  2010. 
 The spectra were re-binned  to 
have a minimum of 20 counts per energy  bin, and energy 
channels below 0.3 keV and above 10 keV were excluded from the  analysis. 
 The XSPEC ({\it v12.5.1}) software was utilised for the analysis. 
 \\
 We extracted a spectrum of the complete first observation (obsid 00332931000) from about 2.7 ks to about 51 ks for a total XRT exposure time of 21.6 ks. We fitted the spectrum  with a model composed of a power-law with two absorption components at low X-ray energies, \texttt{wabs} and \texttt{zwabs}. 
The first one corresponds to  Galactic absorption and its column density  is fixed to the Galactic value $N_{\rm H}^{\rm gal}=2.8\times 10^{20}{\rm cm}^{-2}$ (from Kalberla et al.  2005).
The second absorption is due to the material located at the redshift of the source and its column density
$N_{\rm H}^{\rm host}$  was left free to vary.  The 90\% confidence intervals on the best-fit parameters are obtained with the  \texttt{error} command in XSPEC. This spectrum is well fitted by the single power-law model with a $\chi^2_{red}=1.06$ for 69 dof. The best-fit value of the  host galaxy absorption component is $N_{\rm H}^{\rm host}=5.3^{+6.2}_{-4.3}\times 10^{21}{\rm cm}^{-2}$ and the spectral index is $\beta_{\rm X}=0.96\pm 0.09$ (where the standard notation $F_{\nu}\propto \nu^{-\beta}$ is used).  In order to test for possible spectral evolution, we divided the first observation into two time intervals. The first one corresponds to the first two orbits before 11 ks after the trigger and the second covers the rest of the observation. The best-fit parameters of both spectra are consistent  with each other. Results of the X-ray spectral analysis can be found in Table 1.
\begin{table}
\begin{center}
\begin{tabular}{llllllll}
\hline
\hline
$t_{\rm start}-t_{\rm end}$&$\beta_{\rm X}$ &$N_{\rm H}^{\rm host}$  &$\chi^2_{\rm red}$  \\ 
     s after trigger            &
&  $10^{21}$ cm$^{-2}$   &      \\ 
\hline
\hline
$(0.27$-$5.1)\times 10^4$ & $ 0.96^{+0.09}_{-0.09}$ & $5.3^{+6.2}_{-4.3}$ & 1.06\\
$(0.27$-$1.1)\times 10^4$ & $ 0.93^{+0.12}_{-0.11}$ & $4.0^{+8.1}_{-4.0}$ & 0.65\\
$(1.4$-$5.1)\times 10^4$ & $ 1.02^{+0.16}_{-0.17}$ & $7.6^{+9.9}_{-7.6}$ & 1.36\\
\hline
\hline
\end{tabular}
\label{xrt}
\caption{Results of the X-ray spectral fitting.  We report
 the time interval in which  the
  spectrum was extracted, the unabsorbed spectral index $\beta_{\rm X}$,  the equivalent neutral hydrogen
  column density at the host redshift  $N_{\rm H}^{\rm host}$, the reduced $\chi^2$.
  }
\end{center}
\end{table}   

\section{Afterglow temporal evolution}
\label{lc}
\subsection{Optical and near-infrared light-curve}
\begin{figure}
\vskip -0.7cm
\hskip -9 cm
\psfig{figure=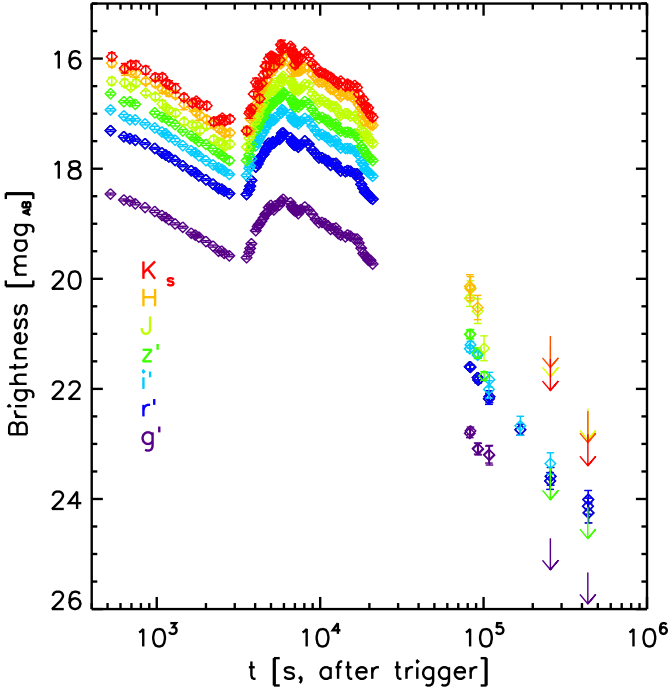,angle=0,width=10cm}
\caption{Observed GROND seven-band light-curve  of the afterglow of GRB 081029. 
Fluxes were not  corrected for Galactic foreground extinction. The full GROND data set is available as on-line material.
}
\label{lcopt}
\end{figure} 

GRB 081029 is characterised by  a complex optical and near-IR light-curve as shown in Fig. \ref{lcopt}. We can divide the optical-NIR light-curve into three phases:
\begin{itemize}
\item[i)] from $10^2$ to  $3.5\times 10^3$ s: a curved shallow decay phase; 
\item[ii)] from  $3.5\times 10^3$ to   $2\times 10^4$ s: a sudden rebrightening of about 1.1 magnitudes in all GROND bands followed by a shallow decay phase superposed with at least three small additional peaks;
\item[iii)] from $2\times 10^4$ to $4.4\times 10^5$ s: a steep decay phase.
\end{itemize}
In the following section we will describe  the observed optical-NIR evolution during these three phases.

\subsubsection{Phase i)}
 The seven GROND  optical-NIR  light-curves of phase i)  can be well represented by a smoothly-connected broken power-law ($\chi^2_{\rm red}=0.95$) while a single power-law fit is excluded  ($\chi^2_{\rm red}=4.4$).  If we contemporaneously fit these seven bands using the parametrisation form Beuermann et al. (1999)
\begin{equation}
F^{(i)}(t)\propto \left[ \left( \frac{t}{t^{(i)}} \right)^{s^{(i)}\alpha_1^{(i)}}+ \left( \frac{t}{t_{{\rm b}}^{(i)}} \right)^{s^{(i)}\alpha_2^{(i)}} \right]^{-\frac{1}{s^{(i)}}},
\label{broken}
\end{equation}
where $\alpha_1^{(i)}$ ($\alpha_2^{(i)}$) is the pre-(post-)break decay index\footnote{Note that in this equation we use the standard notation in which positive (negative) values of $\alpha$ imply a decline (rise) of the light-curve.},  $s^{(i)}$ is the sharpness of the break, the apex $(i)$ the light-curve phase we are describing and the break time  ${t_{{\rm b}}^{(i)}}$ is defined as,
$$t_{\rm b}^{(i)}=t^{(i)}\left(-\frac{\alpha_1^{(i)}}{\alpha_2^{(i)}}\right)^{\frac{1}{s^{(i)}\left(\alpha_2^{(i)}-\alpha_1^{(i)}\right)}}.$$
 We obtain $\alpha_1^{(i)}=0.38 \pm 0.05$ and $\alpha_2^{(i)}=1.12 \pm 0.06$, and an achromatic break  ${t_{{\rm b}}^{(i)}}$  located at 940 $\pm$ 30 s. The optical magnitudes observed by the REM telescope 2.5 min after the trigger (Covino et al. 2008a) are consistent with the extrapolation to earlier times of this curve. A significant deviation from this model can be seen after 2.2 ks, immediately before the start of the intense rebrightening, where the light-curve commences a flattening. Unfortunately, we lack observations between 2.7 ks and 3.5 ks, exactly around the beginning of the rebrightening.

\subsubsection{Phase ii)}
\begin{figure}
\vskip -0.7cm
\hskip -9 cm
\psfig{figure=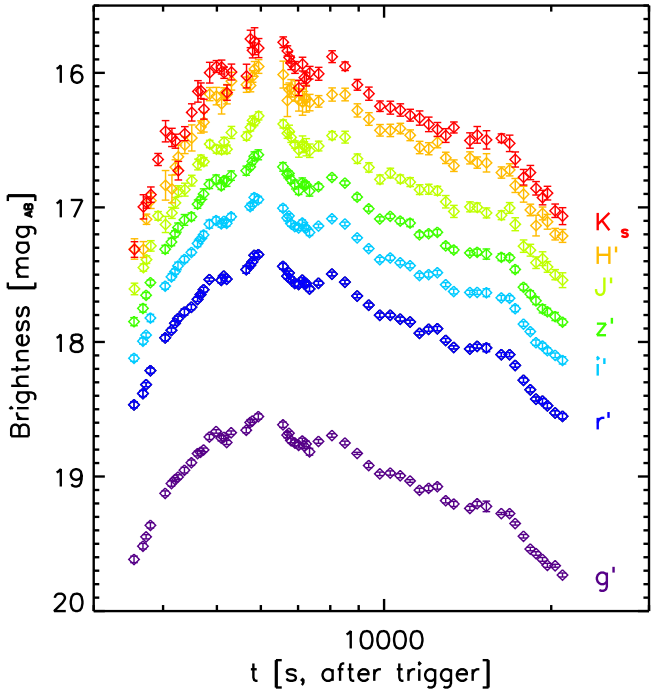,angle=0,width=10cm}
\caption{Observed GROND seven-band light-curve of the afterglow of  GRB 081029 during phase ii).
}
\label{zoom}
\end{figure} 
A steep rise is observed in all seven GROND bands and starts between 2.5 ks and 3.5\,ks. The lack of observations during this interval does not allow us to precisely test the achromaticity of this start. Between 3.5 ks and 4.8 ks the light-curve brightens in all bands by more than 1.1 mag. This rise is very well tracked by the GROND photometry with twelve 1-min observations in all  seven bands.  After a short constant flux state lasting about 400 s, around 5.2 ks after the trigger another rise of about 0.2 mag leads to the maximum at 5.9 ks (see Fig. \ref{modeltriple}). In this discussion we consider this further steep rebrightening at 5.2 ks as the first of a series of optical flares superposed on the post-break shallower power-law continuum. The brightness of the afterglow and the consequent small error bars of the optical photometry  during the rise allow the identification of several  substructures  that make the light-curve deviate from a simple power-law. Considering only the data after 3000\,s, the steep rise requires a power-law index $\alpha_1^{(ii),(iii)}=-4.7$ and a break around 4500\,s. The position of the break is not well constrained because  of the ``flare'' at 5.2\,ks. If we take into account the possible contribution of the broken power-law component we used while fitting the phase i) light-curve, the  steepness of the rise is even more extreme. 

The phase ii) light-curve after the break  is characterised  by an intense variability (see Fig. \ref{zoom}). Fig. \ref{modeltriple} shows that it can be reproduced by a curved continuum with at least three flares superposed at $t\sim5.9$, 8.1 and 18\,ks with a similar  flux excess with respect to the continuum ($\sim$ 0.16 magnitudes in the $r^\prime$ band) and similar logarithmic duration ($\Delta t\sim$ 0.05 dex). Another two possible less intense events ($\Delta$ mag$\sim$0.07 magnitudes in the $r^\prime$ band) can also be seen at $t\sim10$\,ks and $t\sim12$\,ks. The flares are visible in all seven bands and the peak-times are contemporaneous within errors. The continuum shallow decay underlying the flaring activity can be well described by a temporal index $\alpha_{\rm shallow}=0.47$.
\subsubsection{Phase iii)}
\label{phase3}
\begin{figure}
\vskip -0.7cm
\hskip -9 cm
\psfig{figure=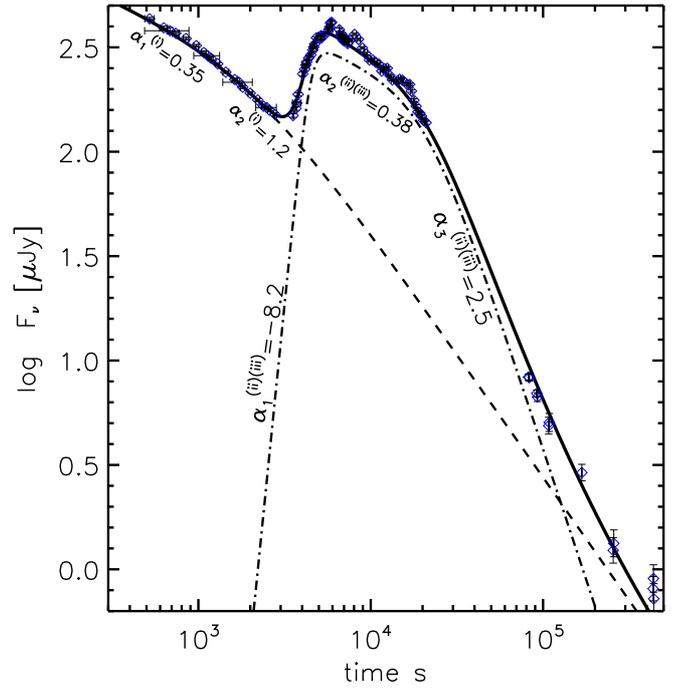,angle=0,width=10cm}
\caption{$r^\prime$ band GROND light-curve of the afterglow of GRB 081029 modelled as a superposition of two separate components as discussed in \S \ref{lcfull}.
}
\label{modeltriple}
\end{figure} 
After a break, the optical-NIR light-curve  steepens at around 19 ks.  Unfortunately, we lack information about the light-curve decay between 21 ks and 82 ks, owing to the day-time constraint in La Silla. During this unobserved window the light-curve  declined by 3 mag, and the GROND observations obtained during the second night (between 82 ks and 108 ks after the trigger) seem to be well connected to the later observations of phase ii) by a unique power-law with a temporal index of about 2.3. Observations in subsequent nights, though, are brighter than the extrapolation of this steep power-law, indicating a flattening of the light-curve after about two days. 
\\
 The complete light-curve of phases ii) and iii) can be well fitted with a smoothly-connected triple power-law when excluding from the fit the flaring activity observed during phase ii) and the late ($t>2\,d$) data (see Fig. \ref{modeltriple}). Any simpler model (i.e., with a smaller number of free parameters such as e.g., a smoothly-connected double power-law) is excluded ($\chi^2_{\rm red}\gg10$).
Using for the smoothly-connected triple power-law a similar functional form as in Eq. \ref{broken},  we obtain for phase  ii) and iii) 
$\alpha_1^{(ii),(iii)}=-4.5\pm 0.4$, $\alpha_2^{(ii),(iii)}=0.47\pm 0.03$, and $\alpha_3^{(ii),(iii)}=2.3\pm 0.2$ with a $\chi^2_{\rm red}=1.15$. The late-time ($t>2\,d$) fluxes are underestimated by the extrapolation of this model.

\subsubsection{The complete GROND light-curve}
\label{lcfull}
We finally tried to fit the whole light-curve with an empirical model consisting of the sum of a smoothly broken power-law such as Eq. \ref{broken} and a  smoothly-connected double power-law model. The result is similar to what we reported before, except that the flux contribution from the double power-law component makes the required temporal index of the rising phase even more extreme   $\alpha_1^{(ii),(iii)}= -8.2$. For the same reason the other decay phase indices are slightly steeper than the ones reported above. Using the  same formalism as before, we obtain  $\alpha_1^{(i)}=0.351 \pm 0.05$, $\alpha_2^{(i)}=1.2 \pm 0.07$,  $\alpha_1^{(ii),(iii)}=-8.2\pm 0.4$, $\alpha_2^{(ii),(iii)}=0.38\pm 0.05$,  and $\alpha_3^{(ii),(iii)}=2.5\pm 0.25$  with a $\chi^2_{\rm red}=1.5$ (obtained excluding the small sub-flares from the fit). This model is the one reported in Fig. \ref{modeltriple}. The slightly higher value of $\chi^2_{\rm red}$ is mainly because of the difficulty in reproducing the sharp transition around 3.5 ks and the possible light-curve flattening after two days.  The temporal breaks are consistent within errors with the separate fits. In this complete light-curve modelling, the contribution of the broken power-law component is  important also at late  times ($t>1 \times 10^5$\,s), decreasing the inconsistency of late observations with the triple power-law model. A residual excess is, however, still present, suggesting either the presence of an underlying dim host galaxy or a further change in the optical decay index. If the broken power-law component were affecting the late-time light-curve, a contribution from  $r^\prime$=25 mag host galaxy would account tor the brighter photometry. When such a component is not contributing at late time, a brighter $r^\prime$=24.6 mag host is required. As discussed in section \ref{grondobs}, we can exclude the presence of an underlying host galaxy down to a limiting magnitude of $r^\prime_{\rm host}=25.8$ and therefore this flattening is likely related to the afterglow evolution.

\subsection{X-ray light-curve}
\begin{figure}
\vskip -0.7cm
\hskip -9 cm
\psfig{figure=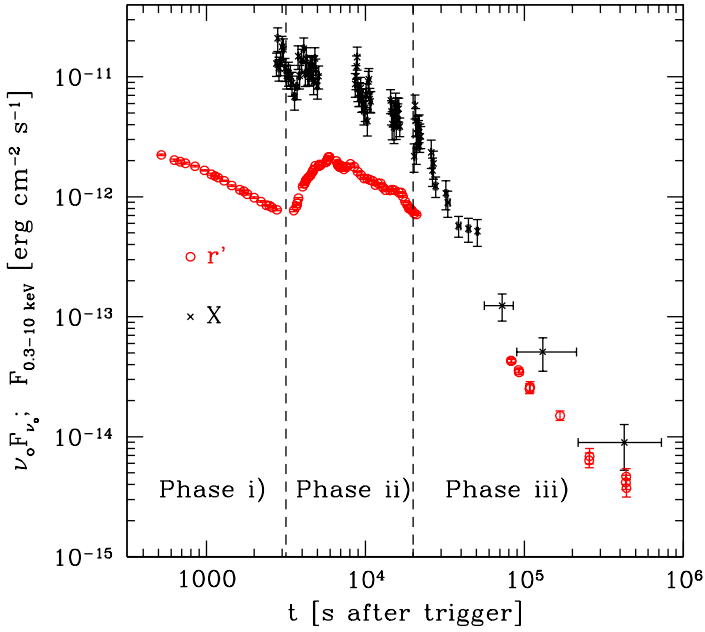,angle=0,width=10cm}
\caption{$r^{\prime}$ band $\nu F_{\nu}$ light-curve (empty circles) superposed on the unabsorbed 0.3--10 keV XRT light-curve (black crosses) of the afterglow of GRB 081029. The $r^{\prime}$ band data were corrected for the Galactic foreground  extinction.  Vertical dashed lines represent the transition time between the different phases described in \S \ref{lc}.
}
\label{lcobs}
\end{figure} 
Owing to observing constraints (Sakamoto et al., 2008), XRT started to follow-up  GRB 081029 only 2.7 ks after the trigger. Observations were performed  in PC mode and found an X-ray counterpart with a 0.3 -10 keV count rate of about 0.3 cts, corresponding to an unabsorbed flux of about 1.3$\times 10^{-11}$ erg cm$^{-2}$s$^{-1}$ following the spectral analysis reported in section \ref{xspectrum}. The count rate of the XRT light-curve was downloaded from the UK \swift Science Data Centre\footnote{\tt http://www.swift.ac.uk/xrt\_curves/} (see Evans et al. 2007, 2009 for an extended description of the data reduction). The X-ray light-curve does not show any evidence of a rebrightening contemporaneous to the one observed by GROND (see Fig. \ref{lcobs})  and  can be described by a broken power-law with temporal indices $\alpha_{1, X}=0.48\pm 0.1$ and $\alpha_{2, X}=2.4\pm 0.17$ ($\chi^2_{\rm red}=1.13$). A single power-law model is excluded ($\chi^2_{\rm red}=4.9$). A  small fluctuation ($\Delta F\sim 0.1$ dex) from a straight power-law evolution is visible during the first orbit (see \S \ref{discussion}).  These temporal indices are consistent with the values obtained for the optical-NIR evolution after the bump ($\alpha_2^{(ii),(iii)}=0.47\pm 0.03$, $\alpha_3^{(ii),(iii)}=2.3\pm 0.2$). The temporal break is located at $1.8^{+1.5}_{-0.9}$\,ks, in perfect agreement with the break of the second component in the  GROND light-curve.

\section{Colour evolution and spectral energy distribution (SED)}
\label{sed}

\begin{figure}
\vskip -0.7cm
\hskip -9 cm
\psfig{figure=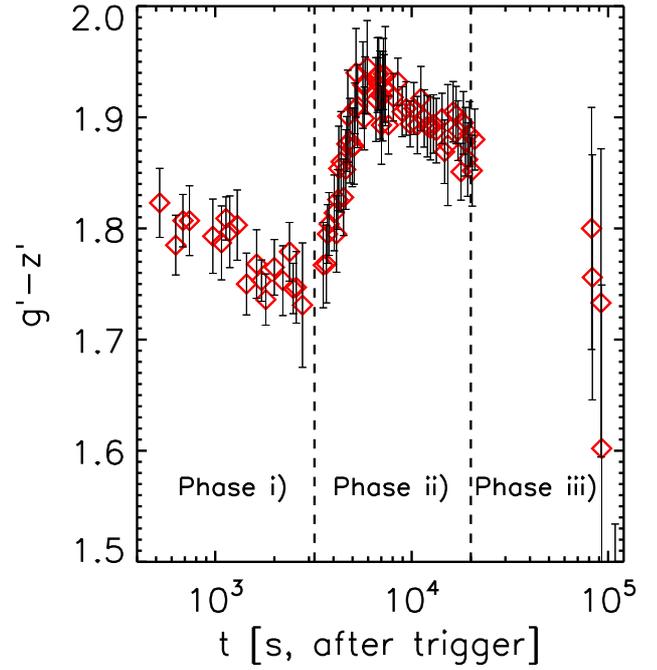,angle=0,width=10cm}
\caption{Temporal evolution of the  AB magnitudes differences between GROND  $g^\prime$ and  $z^\prime$ bands.   Vertical dashed lines represent the transition time between the different phases described in \S \ref{lc}.
}
\label{g_z}
\end{figure} 
\begin{figure}
\vskip -0.7cm
\hskip -9 cm
\psfig{figure=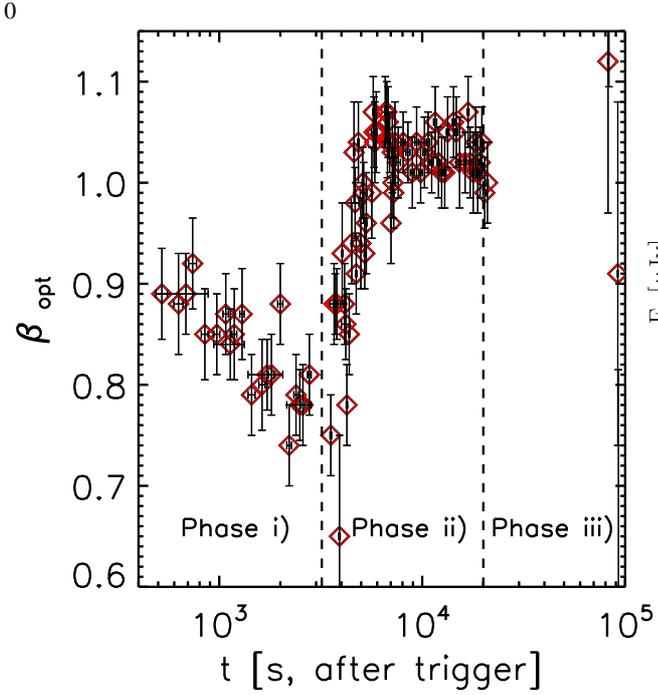,angle=0,width=10cm}
\caption{Temporal evolution of the optical spectral index $\beta$ obtained fitting the  unabsorbed  $K_s, H, J, z^\prime, i^\prime, r^\prime$ GROND band SED for all the available observations as discussed in \S \ref{sed}.  Vertical dashed lines represent the transition time between the different phases described in \S \ref{lc}.
}
\label{lcbeta}
\end{figure}

Thanks to GROND's capability of obtaining images in seven bands contemporaneously,  it was possible to study the colour evolution during a bright optical rebrightening without requiring any temporal extrapolation. Thanks to the very small errors in the GROND photometry due to the good sky conditions during the first night of observations and thanks to the brightness of the source, we are able to study the evolution of the  colour between different bands in every exposure. Comparing the magnitudes observed in different bands we  clearly see a sudden reddening during the optical rebrightening (see Fig \ref{g_z}). A less prominent colour evolution is observed during phase i) with the $g^\prime-z^\prime$  colour getting bluer by about $\Delta {\rm mag}\approx 0.1$ between 400\,s and 3000\,s. We cannot exclude a further less prominent colour evolution during  phase iii)  while the hint of colour evolution during the small rebrightenings observed during phase iii) is not statistically significant. 
\\
 In order to estimate the possible effect of the host galaxy dust absorption, we extracted the optical-NIR SED of GRB 081029 at different times before and after the bump.  GRB 081029 occurred at z=3.8479, therefore both the GROND $g^\prime$ (and partially the $r^\prime$) bands are affected by the Lyman alpha absorption. Because of  the uncertain intergalactic hydrogen column density along the line of sight, the $g^\prime$ band is excluded from the SED fits. We fitted  the other six optical-NIR GROND bands (i.e., $K_s, H, J, z^\prime, i^\prime, r^\prime$) assuming a simple power-law spectrum after correcting the observed fluxes for the  foreground Galactic extinction of  $E_{\rm B-V}=0.03$ mag (Schlegel et al. 1998)  corresponding to an extinction of $A_{\rm V}^{\rm Gal}= 0.093$ mag using $R_{\rm V}$ = 3.1. Large and Small Magellanic
Clouds (LMC, SMC) and Milky Way (MW) extinction laws from Pei (1992) were used to describe the dust reddening in the host galaxy. We found that all  SEDs are consistent with a negligible host galaxy dust absorption for all considered extinction curves. Using a SMC extinction curve, we obtained a 90\% confidence level upper limit for the host galaxy extinction $A_{\rm V}^{\rm host}< 0.16$ mag at 10.9\,ks.  The $K_s, H, J, z^\prime, i^\prime, r^\prime$ spectral index (where the standard notation $f(\nu)\propto \nu^{-\beta}$ is adopted) is $\beta_{\rm opt}=1.06^{+0.06}_{-0.05}$ and the reduced $\chi^2$ is 1.06.  In Fig. \ref{lcbeta} we show the temporal evolution of the optical spectral index  for each individual GROND observation. The colour evolution during the light-curve bump shown in Fig. \ref{g_z}  is also clearly visible in Fig. \ref{lcbeta} where the spectral hardening is also observed before the bump.  This latter early-time colour evolution is consistent with the light-curve break observed around 900\,s to be caused by a spectral break moving bluewards through the observed GROND bands. The observed $\Delta \beta_{\rm opt}$ could be related to an incomplete passage of a cooling frequency in a wind-like profile if the shape of the cooling break were very shallow. The slow decay rates observed during phase i) do not agree with the standard closure relations when compared with the values of $\beta_{\rm opt}$ (see, e.g., Racusin et al. 2009) which cast some doubts on this interpretation. The nature of the optical-NIR afterglow reddening during the rebrightening will be discussed in \S \ref{model}.

\begin{figure}
\vskip -0.0cm
\hskip -9 cm
\psfig{figure=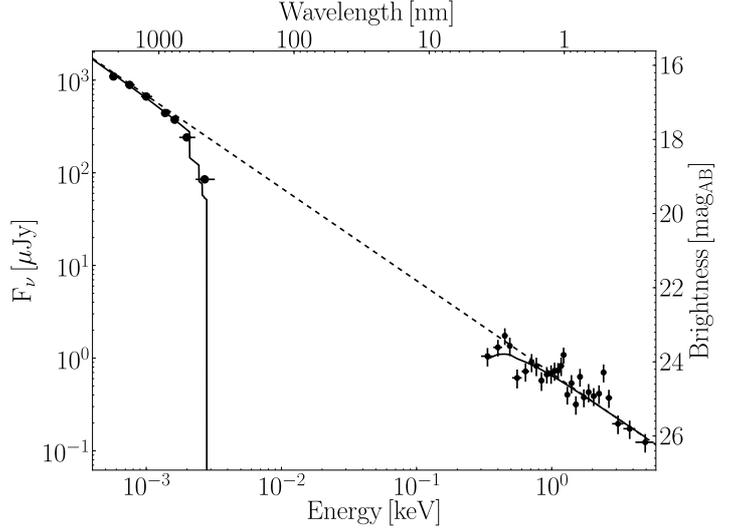,angle=0,width=9.5cm}
\caption{Combined GROND and XRT spectral energy distribution extracted around 11ks as discussed in \S \ref{sed}. The dashed line represents the unabsorbed single power-law connecting the X-rays with the NIR bands and characterised by a spectral index $\beta=1.00\pm 0.01$. The solid line represents the same absorbed model with $A_{\rm V}^{\rm host}=0.03^{+0.02}_{-0.03}$ mag and $N_{\rm H}^{\rm host}=7.9^{+6.8}_{-5.9}\times 10^{21}~{\rm cm}^{-2}$. Note that the $g^\prime$ (and partially the $r^\prime$) bands are strongly affected by the Lyman break at z=3.8.
} 
\label{sedoptx}
\end{figure}

Because the optical and X-ray light-curves show a similar temporal evolution after the bump,  we can analyse the combined optical to X-rays spectral energy distribution under the simple assumption that the flux observed in both bands is produced by the same mechanism. Following the method described in Greiner et al. (2011), we extracted a broad-band SED  around 10\,ks. We selected this time interval to obtain a contemporaneous GROND and XRT coverage and to avoid the presence of the small sub-flares observed during the rebrightening.  We find that the SED is well fit by a  single power-law connecting the NIR to the X-ray band (see Fig. \ref{sedoptx}). Moreover, no host galaxy dust absorption is required in this case.  The best-fit value for rest-frame reddening is  $A_{\rm V}^{\rm host}=0.03^{+0.02}_{-0.03}$ mag, the broad-band spectral index is  $\beta=1.00\pm 0.01$ with a host galaxy column density $N_{\rm H}^{\rm host}=7.9^{+6.8}_{-5.9}\times 10^{21}~{\rm cm}^{-2}$, which is consistent with both $\beta_{\rm opt}$ and  the values of $\beta_{\rm X}$ and $N_{\rm H}^{\rm host}$ from the XRT  spectral analysis reported in Tab. 1. 

\section{Nature of the optical rebrightening}
\label{model}
The most evident peculiarity of GRB 081029 is the presence of the intense optical rebrightening that occurs around 3.5 ks after the trigger. This is not the first case in which a late-time optical rebrighening is observed in  a long GRB afterglow but the intensity of the flux increase and  the steepness of the light-curve rise are unusual. Moreover,  the density of the available data set and the possibility to contemporaneously study the temporal evolution of the rebrightening in seven bands thanks to GROND make GRB 081029 a unique case.
\\
A sudden increase of optical flux was already observed in one of the first afterglows ever detected. GRB 970508 showed a rebrightening of about 1.5 magnitudes in I$_c$, R and V bands around one day after the trigger (Sahu et al. 1997; Vietri 1998; Sokolov et al. 1998; Nardini et al. 2006). Other cases reported in the literature are GRB 060206 (Monfardini et al. 2006) , GRB 070311 (Guidorzi et al. 2007; Kong et al. 2010), GRB 071003 (Perley et al. 2008; Ghisellini et al. 2009), and GRB 071010A (Covino et al. 2008b; Kong et al. 2010).
\\
Several interpretations have been proposed for explaining the rebrightenings of these different GRBs but none of them is able to take into account all the  characteristics shown by the different light-curves under a unique physical framework. In this section we give a brief overview of the different proposed models and  will then check if  the broad-band light-curve of GRB 081029 can be explained in the framework of some of these models.

\subsection{Discontinuity  in the density profile}
In the external shock  model, a sudden increase of the external medium density can, in principle, produce a rebrightening in the observed afterglow light-curve  (Lazzati et al. 2002). 
Such a density profile can be found in the surroundings of a long GRB. This is caused by the impact of the stellar wind on the interstellar medium  (Kong et al. 2010).
This  effect should be prominent  for frequencies between the typical synchrotron frequency and the cooling frequency $\nu_{\rm m}<\nu<\nu_{\rm c}$,  therefore this could explain the presence of a rebrightening in the optical without a corresponding flux increase in the X-rays if the cooling frequency is located between those bands. The presence of density-jumps  was used  to explain the fluctuations in the optical light-curves of GRB 030226 (Dai \& Wu 2003) and  GRB 021004  (Lazzati et al. 2002; Nakar \& Piran 2003).
However, a change in the external density profile is not expected to produce a colour evolution in the optical-NIR bands  if the condition  $\nu_{\rm m}<\nu<\nu_{\rm c}$ persists. Because the cooling frequency is expected to move redwards with increasing density (see, e.g., Panaitescu and Kumar 2000), we would then require the density jump to be high enough to allow  $\nu_{\rm c}$  to move redwards of the observed $K_s$ band to reproduce the observed colour evolution.
\\
However, numerical simulations have shown that even sharp discontinuities  with a strong increase in the density profile of a uniform external medium, or the encounter of the blast wave with a wind termination shock, cannot produce a sharp bump in the observed light-curve and only a smooth and diluted change can be detected (Nakar and Granot 2007; van Eerten et al. 2009).
For this reason,  a simple change in the external medium density is not a reliable explanation because the intense rebrightening of the optical-NIR light-curve of GRB 081029 is very sharp.

\subsection{Variation of micro-physics in wind bubbles}
In a recent paper, Kong et al. (2010), tried to solve the problem of reproducing sharp light-curve changes for a jump in the density profile. They divide the circum-burst environment into two regions: the first one characterised by a stellar wind density profile, the second one by a shocked stellar wind mixed with a small
fraction of the swept-up ISM.  Letting the micro-physical parameters $\epsilon_{\rm e}$ and $\epsilon_{\rm B}$  vary with the electron energy $\gamma$ in a different way in the two regions, their synthetic light-curves are able to reproduce the sharp optical rebrightenings observed in some GRBs. In their analysis,   Kong et al. (2010) studied only GRBs with contemporaneous bumps in the optical and in the X-rays and  took into account only a single optical band without checking for possible colour evolution. For  GRB 081029,  the X-ray light-curve does not show any contemporaneous rising phase and we also observe a clear colour evolution in the GROND energy range during the bump. A fine-tuning in the assumed evolution of the micro-physical parameters would  therefore be required  to force $\nu_{\rm c}$ to pass through the optical-NIR energy range at exactly the same time as the changing of the density profile to observe a similar colour evolution. The analysis of the effects of a variation of  $\epsilon_e$ and $\epsilon_{\rm B}$ on the cooling frequency is required to test if this scenario can also explain cases like GRB 081029.  Also in this case, however, the pre-bump evolution observed in the X-rays lacks explanation unless the X-ray frequencies are completely unaffected by these changes.

\subsection{Late prompt model}
\label{late}
In the late prompt model (Ghisellini et al. 2007, 2009; Nardini et al. 2010), the observed optical, NIR, and  X-ray light-curves of GRBs are produced by the sum of two separate components: the standard forward shock afterglow emission and a radiation related to a late-time activity of the central engine sustained by the accretion of the fall-back material that failed to reach the escape velocity of the progenitor star. The combined effect of these two components can reproduce a large variety of observed light-curves (Ghisellini et al. 2009) and is also used to describe the presence of late-time optical rebrightenings in the light-curves of GRB 071003 (Ghisellini et al. 2009) and  GRB 061126 (Ghisellini et al.  2009; Nardini et al. 2010). In these events, the excess of optical flux with respect to the extrapolation of the earlier light-curve appears because the standard afterglow emission is decaying as a power-law while the late prompt component is still  constant or slightly increasing and therefore it becomes dominant at late times.  Because the two components  originate in completely different mechanisms, a colour evolution in the optical-NIR SED during the bump is expected, similar to the one obtained for GRB 081029. A colour change related to an increasingly dominant second component is indeed observed both in GRB 071003 and in GRB 061126 (Ghisellini et al.  2009, Nardini et al. 2010).   On the other hand, in the late prompt model (Ghisellini et al. 2007), the early-time flat (or slowly increasing) evolution  of the late prompt component is caused by a geometrical effect (i.e., increase of the visible emitting surface owing to a decrease of the $\Gamma$ lorentz factor of the slower ``late prompt'' shells). For GRB 081029 the rise of the optical rebrightening is extremely steep which renders this scenario unlikely unless a delayed reactivation of the central engine is responsible for the emission of these less energetic shells. This delayed starting point would allow a new definition of the late prompt component starting time and  consequently a flattening of this component's light-curve rise.

\subsection{Two-component jet Model}
\label{2jets}
Another scenario that invokes the coexistence of two  separate emitting processes, and therefore allows the observation of a discontinuity in the light-curve  with a contemporaneous colour evolution, is the two-component jet model. The presence of a fast narrow jet, co-alligned with a wider, slower jet has been claimed to be able to describe the complex broad-band evolution of some GRBs (e.g., GRB 030329 Berger et al.; 2003, GRB 080319B Racusin et al. 2009; GRB 080413B Filgas et al. 2010). In this scenario, a flattening or a bump in the observed light-curve can be observed when the onset of the afterglow produced by the slower (and wider)  jet occurs when the afterglow emission of the faster (and narrower) jet is already decreasing fast because of an early-time jet break. The rising wider jet afterglow can be characterised by a different SED in the observed bands and therefore a colour evolution can be observed during the transition. At later times the observed radiation is dominated by the wider jet afterglow.  
\\
The complexity  of the GRB 081029 light-curve is hard to  explain in the framework of this scenario. The flat early-time optical light-curve observed before the rebrightening, when interpreted in the framework of a standard forward shock afterglow scenario, suggests an environment characterised by a wind-like density profile. On the other hand, to obtain a steep rebrighening at later times, the second jet afterglow light-curve requires an ISM-like external medium. Moreover, the optical rise that is required for two separate components is much steeper than the pre-peak rise previously observed in other GRBs (Rykoff et al. 2009; Panaitescu \& Vestrand 2008). In the standard forward shock emission scenario for an isotropic outflow seen face-on by the observer, during the pre-deceleration phase the optical light-curve is supposed to rise as $F\propto t^2$ or $t^3$ depending on the location of the cooling frequency.  A steeper  $F\propto t^4$ rise can be obtained for of an off-axis observer for a collimated outflow with a sharp angular boundary (Panaitescu \& Vestrand 2008). No steeper rise, similar to the one observed for GRB 081029 can be reproduced under these standard assumptions. Panaitescu \& Vestrand (2008) also found that the afterglows showing a fast pre-peak rise ($\alpha < -1$) are characterised by a correlation between the $k$-corrected optical flux  at the peak and the redshift-corrected  peak time assuming a common redshift $z=2$. Considering only the contribution of the second component corrected for the Galactic foreground absorption (i.e., the triple power-law described in section \ref{lc}), the  peak flux calculated in the r-band re-locating GRB 081029 at $z=2$ is $F_{\rm p}^{z=2}=1.3$ mJy. The peak time at redshift 2 is $t_{\rm p}^{\rm z=2}=3.5$ ks. Comparing these values with the plot shown in Fig. 2 of Panaitescu \& Vestrand (2008), the rebrightening of GRB 081029 is inconsistent with this correlation with GRB 081029 having a peak flux that is much brighter than expected for the peak time. That confirms the difficulties of explaining the fast optical bump of GRB 081029 as the signature of a second jet afterglow with the same initial time as the one responsible for the early-time light-curve.

\subsection{Energy injection}
An alternative scenario invoked for explaining late-time optical bumps is  a discrete episode of energy injection into the fireball by the late-time interaction of slow shells with the forward shock (J\'ohannesson et al. 2006;  Fan \& Piran 2006; Covino et al. 2008b; Rossi et al., 2011).  In  GRB 081029, the steepness of the optical bump and the large ratio between the  second component and the extrapolation of the early broken power-law decay (between $\sim$ 5 and 8 during the bump) imply a sudden  release of a large amount of energy at late times.  Such an energy injection is not supposed to produce a change in the observed synchrotron spectral slopes under the standard assumption of non-evolving micro-physical parameters in the fireball (e.g., $\epsilon_{\rm e}$, $p$,  $\epsilon_{\rm b}$) Under this assumption, an observed spectral evolution can only be produced if  one of the characteristic frequencies (e.g., the cooling frequency $\nu_c$) crosses the observed bands. On the other hand, we know that no contemporaneous  rebrightening is observed in the X-rays. This evidence acts against this scenario because an episode of energy injection would increase the normalisation of the whole synchrotron spectrum, which would be visible at all wavelengths.

 \section{Discussion}
 \label{discussion}
The complex light-curve of GRB 081029 is a remarkable example of how the increasing quality of optical-NIR follow-up represents a hard test for the proposed afterglow emission models. In \S \ref{model} we have shown how the most commonly invoked extensions of the standard external shock afterglow models fail to reproduce the steep optical rise observed in the GRB 081029 light-curve without invoking fine-tuned sets of parameters that still lack a convincing physical interpretation. The component that is dominating the early-time pre-bump optical-NIR light-curve is likely still present, although weak,  during the optical bump and becomes dominant again after $10^5$\,s. This evidence, together with the sudden colour evolution that accompanies the optical rebrightening, favours the two-component nature of this light-curve.  As discussed in sections \ref{late} and \ref{2jets}, both the late prompt an the two-component jet models fail to explain the extremely steep rise of the second component ($F(t)\propto t^{\sim 8}$) in their standard formulation.  This problem could be solved by re-scaling the time at which  the second component starts to act. If we shift  the starting point of the second component $t_{0,2}$ from the gamma-ray detector trigger time to the mean time of the first GROND observation in which we detect the rebrightening (i.e. $t_{0,2}=3500$\,s after trigger), we have the shallowest value of the rise we can obtain with this simple temporal rescaling.  Taking into account the contribution of the ongoing early-time component, we obtain a rising temporal index of about  $\alpha_{1,{\rm rescaled}}^{(ii),(iii)}\approx -0.7$ for the second component. We can therefore obtain any value for $\alpha_{1,{\rm rescaled}}^{(ii),(iii)}$ between -0.7 and -8.2 selecting a new $t_{0,2}$ between the trigger time and 3.5 ks. For example, a slope  $\alpha_{1,{\rm rescaled}}^{(ii),(iii)}\approx -2.5$, typical  of the {\it fast optical rise} afterglows studied by Panaitescu \& Vestrand (2008), can be obtained by shifting the initial time to $t_{0,2}\approx 2500$\,s after trigger. Using this new observer frame $t_{0,2}\approx 2500$\,s,  the peak-time at redshift 2 is $t_{\rm p}^{\rm z=2}=2.0$ ks. This value decreases  the inconsistency with the correlation found by Panaitescu \& Vestrand (2008), but the peak of the second component of GRB 081029 still lies above the correlation reported in that paper.  
\\
Rescaling the $t_{0,2}$ requires a reactivation of the central engine $\sim$ 0.5 ks  (2.5 ks in the observer-frame) after the main event. Such a rescaling of  $t_{0,2}$ implies that a new process separated from the one producing the early optical light-curve is activated several hundred seconds after the main event. If this process is related to a delayed reactivation of the central engine, this could produce an observable signature at higher frequencies around the new  $t_{0,2}$. The existence of prompt GRB light-curves showing long periods of quiescence between separate peaks lasting up to a few hundreds seconds has already been discussed in the literature (Ramirez-Ruiz, Merloni \& Rees,  2001; Romano et al. 2006; Burlon et al., 2008, 2009; Gruber et al. 2010). Most of them are pre-cursors (i.e., peaks preceding the main GRB event), but Burlon et al. (2008) report on the case  of GRB 060210 where two post-cursors are observed 60 and 150s after the end of the main burst. There is no clear example with later time ($t \geq 1500$ s) post-cursors, although faint late-time peaks are hardly detectable by BAT (see Holland et al. 2010 for a possible candidate). This late-time activity of the central engine could likely be detectable in the X-rays  as a peak/flare in the XRT light-curve around the new value of $t_{0,2}$. Unfortunately, \swift slewed to GRB 081029 only after 2700 s owing to an observing constraint (Sakamoto et al., 2008), and therefore we cannot constrain the presence of such a signature around 2000-2500 s. However, the {\it Swift} slew was early enough to observe the X-ray afterglow for around 1 ks before the start of the optical rise. This early XRT light-curve remains more or less constant during the first XRT orbit (between 2.7 and 5.5 ks) apart from a small fluctuation of the same order as the flux error bars. No steep rise is therefore shown by the X-ray light-curve simultaneous to the optical one. Because the  X-ray light-curve tracks the GROND one after the optical rise and the broad-band SED is consistent with a single power-law (see section \ref{sed}), we can assume that they are  produced by the same mechanism. Under this assumption the lack of a contemporaneous X-ray rebrightening around 4 ks is barely understandable without invoking an additional flux contribution in the X-rays in the first 1 ks of the XRT observation. The quality of the X-ray light-curve before 3.5 ks is not sufficient to test whether the XRT light-curve before 4\,ks is intrinsically constant or if it is caused by a superposition of a component rising at a later time (similar to the one observed in the GROND bands) with the decreasing tail of an unobserved flare that is occurred before 2.7\,ks. Even if it not possible to constrain a possible X-ray spectral change during these early observations with a spectral fitting, the hardness-ratio light-curve obtained using the publicly available \swift Burst Analiser\footnote{\tt http://www.swift.ac.uk/burst\_analyser/00332931/} (Evans et al. 2010) shows a statistically significant evolution during the first orbit (between 2.7\,ks and 5.1\,ks).  This agrees with the presence of an undetected X-ray flare at earlier times, corresponding with the possible reactivation of the central engine  responsible of the optical and X-ray emission after 4 ks.

\section{Conclusions}
In conclusion, the most commonly invoked extensions of the standard afterglow models fail to explain the steepness of the optical rebrightening observed in GRB 081029 in particular when combined with the associated colour evolution and with the lack of a contemporaneous rise in the X-rays. That the post-rise optical-NIR light-curve tracks the X-ray evolution well suggests the existence of a second component, separated from the one producing the early-time optical afterglow, responsible for both the observed optical and X-rays flux at later times. We are unable to determine the nature of this second component, but from the combined GROND and XRT light-curve we can infer the following phenomenological argument. An easy way to explain the steepness of the rise of this second component is to assume a reactivation of the central engine around 0.5 ks after the prompt emission onset. Unfortunately, the lack of XRT coverage before 2.7\,ks (obs frame) does not allow a direct proof of this hypothesis (e.g., detecting a flare/post-cursor in the X-rays). However, the flux excess in the first 1 ks of the XRT light-curve with respect to the optical-NIR one is consistent with the existence of  an undetected X-ray post-cursor at the  initial time of the second component.


%

\begin{acknowledgements}
We would like to thank the anonymous referee for her/his  useful comments.   
MN acknowledges support by DFG grant
SA 2001/2-1. TK acknowledges
support by the DFG cluster of excellence ``Origin and Structure of
the Universe'', and  ANG, DAK, ARossi, and AU are grateful for travel funding support through MPE.
FOE acknowledges funding of his Ph.D. through the Deutscher Akademischer
Austausch-Dienst (DAAD), SK and ARossi acknowledge support by DFG grant
Kl 766/13-2 and ARossi additionally from the BLANCEFLOR Boncompagni-
Ludovisi, n\'ee Bildt foundation.  Part of the funding for GROND (both hardware as well as personnel)
was generously granted from the Leibniz-Prize to Prof. G. Hasinger (DFG
grant HA 1850/28-1). DB is supported through DLR 50 OR 0405. SK, ANG and DAK acknowledge support by DFG grant Kl 766/16-1. This work made use of data supplied by the UK Swift
Science Data Centre at the University of Leicester.
\end{acknowledgements}

\end{document}